\documentclass[]{aa}
\usepackage{times}
\usepackage{graphics}

\begin{document}


   \title{On OH line formation and oxygen abundances in metal-poor stars}

   \author{Martin Asplund \and Ana Elia Garc\'{\i}a P{\'e}rez
          }

   \offprints{\email{martin@astro.uu.se}}

   \institute{Astronomiska Observatoriet, Box 515, SE-751 20 Uppsala, Sweden \\
              e-mail: martin@astro.uu.se, aegp@astro.uu.se
             }

   \date{{\em Accepted for Astronomy \& Astrophysics}}

\abstract{
The formation of the UV OH spectral lines
has been investigated for a range of stellar parameters
in the light of 3D hydrodynamical model atmospheres. 
The low atmospheric temperatures encountered at low metallicities
compared with the radiative equilibrium values enforced in classical
1D hydrostatic model atmospheres have a profound impact on the OH
line strengths. As a consequence, the derived O abundances using 3D models
are found to be systematically lower by more than 0.6\,dex at
[Fe/H]~$=-3.0$ compared with previous 1D analyses,
casting doubts on the recent claims for a monotonic increase in [O/Fe] towards 
lower metallicities. 
In fact, taken at face value the resulting 3D LTE trend is in rough
agreement with the conventional [O/Fe] plateau. 
Caution must, however, be exercised in view of the remaining assumptions
in the 3D calculations.
We have verified that the stellar parameters remain essentially 
unchanged with 3D model atmospheres
provided that the infrared flux method ($\Delta T_{\rm eff} \la 20$\,K),
Hipparcos parallaxes ($\Delta {\rm log} g \la 0.05$) and Fe\,{\sc ii}
lines ($\Delta [{\rm Fe/H}] \la 0.1$\,dex) are utilised, leaving the
3D O abundances from OH lines largely intact
($\Delta [{\rm O/H}] \la 0.05$\,dex).
Greater concern stems from possible departures from LTE in 
both the line formation and the molecular equilibrium, which, if present,
would increase the derived O abundances again. 
Non-LTE line formation calculations with 1D model atmospheres
suggest no significant steepening of the [O/Fe] trend even if the 
abundance corrections amount to about 0.2\,dex for all 
investigated stellar parameters. We note, however, that
the 3D case may not necessarily be as metallicity-independent.
The apparent lack of laboratory or theoretical 
rate coefficients at the relevant temperatures for the involved 
molecular reactions unfortunately
prevents a quantitative discussion on the possible
effects of non-equilibrium chemistry.
\keywords{Convection -- Line: formation -- Stars: abundances --
Stars: atmospheres -- Stars: Population II -- Galaxy: evolution   }
}

   \maketitle

%

\section{Introduction}

Oxygen is the third most abundant element in the Universe after hydrogen
and helium. Besides being relatively common, oxygen has also attracted a great
deal of attention due to its special role in galactic and stellar evolution.
Since oxygen is predominantly produced in connection with core collapses
of massive stars (type II supernovae, SNe\,II) while iron is also forged 
during accretion-induced collapses of white dwarfs 
(type Ia supernovae, SNe\,Ia),
the abundance ratio of these two elements at different cosmic epochs 
give important insight to the formation and evolution of the Galaxy, as
well as constraining the physics of supernovae. 
The amount of oxygen in stars can also significantly influence the nuclear
energy production and the opacities, which affects stellar evolution.
As a consequence, the dating of globular clusters depend on the stellar
oxygen content: an increase of the oxygen over-abundance 
[O/Fe]\footnote{The abundance ratios are defined by the customary
[X/Fe]=${\rm log} (N_{\rm X}/N_{\rm Fe})_* - 
{\rm log}  (N_{\rm X}/N_{\rm Fe})_\odot$} 
from $+0.3$ to $+1.0$ introduces an about 2\,Gyr
lower age of the oldest clusters
(VandenBerg \& Bell 2001).
Additionally, the light elements Li, Be and B are produced through
cosmic ray spallation of C, N and, most importantly, O with protons
and $\alpha$-particles. Thus, a proper understanding of the oxygen
abundances of stars of different metallicities
is required in order to interpret the evolution of the light elements.

Following the first indication of an oxygen over-abundance relative
to iron ([O/Fe]~$>0$) in metal-poor stars (Conti et al. 1967), 
a great number of
studies have been devoted to quantify this enhancement. Although
all agree on its existence, the amount of the over-abundance is hotly
contested. Various oxygen diagnostics in different types of stars
have been applied but with disparate results. The forbidden [O\,{\sc i}] lines
at 630.0 and 636.3\,nm in metal-poor giants suggest a nearly flat
plateau at [O/Fe]\,$\sim 0.4$ for [Fe/H]\,$\la -1.0$ 
(e.g. Barbuy 1988; Sneden et al. 1991)
while the O\,{\sc i} triplet at 777\,nm in metal-poor dwarfs and subgiants
tend to imply systematically higher values, often with a monotonic
increase towards lower metallicities
(e.g. Abia \& Rebolo 1989; 
Israelian et al. 1998, 2001; Boesgaard et al. 1999; Carretta et al. 2000).  

All oxygen criteria have their pros and cons, which influence the
conclusions. The forbidden lines are immune to departures
from local thermodynamic equilibrium (LTE) 
(cf. discussion in Kiselman 2001)
but the lines are very weak at low metallicities, in particular
in unevolved stars where the feature becomes undetectable for 
[Fe/H]~$\la -2.0$ (Nissen et al. 2001). 
Furthermore, concerns regarding the primordial nature
of the oxygen in field giants have been voiced as mixing of nuclear-processed
material may pollute the surface,
as evident in many globular cluster giants (e.g. Langer et al. 1997).
The O\,{\sc i} triplet on the other hand is more easily discerned 
in metal-poor dwarfs but also more susceptible to departures from LTE 
(cf. Kiselman 2001 and references therein) and inhomogeneities introduced by 
stellar granulation (Kiselman \& Nordlund 1995) than the [O\,{\sc i}] lines. 
The high excitation potential 
of the O\,{\sc i} lines also make them
vulnerable to errors in the effective temperature ($T_{\rm eff}$). 
Additionally, an often overlooked source of confusion is the stellar Fe
abundances, which should be self-consistently computed in order to
obtain reliable [O/Fe] ratios. 
Finally, even the question of the
absolute solar oxygen abundance is still not settled, leaving the 
fundamental reference point for stellar [O/Fe] ratios insecurely anchored. 

An attractive alternative to the O\,{\sc i} and [O\,{\sc i}] lines are
provided by the molecular OH electronic transitions in the UV. 
(Bessell et al. 1984, 1991; Nissen et al. 1994). In fact, at 
[Fe/H]~$\sim -4$ no substitute for the OH lines as the prime oxygen diagnostic 
is available. Therefore the OH spectral
line formation must be properly understood
when attempting to probe the earliest epochs of the Galactic evolution
for example to identify the nucleosynthetic fingerprints 
of the elusive Population III stars (e.g. Karlsson \& Gustafsson 2001).
Recently, Israelian et al. (1998, 2001) and Boesgaard et al. (1999) have
analysed the UV OH (A$^2\Sigma$ - X$^2\Pi$) 
lines in stars down to 
[Fe/H]~$\simeq -3.3$ with high $S/N$ and resolution spectra
and have found a linear trend in [O/Fe] vs [Fe/H] with
a slope of about $-0.4$\footnote{When restricting to the OH-based results
and taking the errors in both [Fe/H] and [O/Fe] into account,
the Boesgaard et al. (1999) data gives a slope of $-0.40\pm0.04$ and 
$-0.36\pm0.04$ for the King (1993) and Carney (1983) $T_{\rm eff}$-scales,
respectively. Similarly, the published results of Israelian et al. (1998)
are consistent with a slope of $-0.38\pm0.07$, which becomes $-0.37\pm0.06$
when also including the new OH data in Israelian et al. (2001).
For simplicity we here adopt a slope of $-0.40$ but emphasize that the exact
value only has a marginal effect on our results. 
It should be noted that the values given in the original references
are slightly smaller due to the combination of OH and O\,{\sc i} results
(Boesgaard et al.) and restriction to stars with [Fe/H]~$\le -1.0$ 
(Israelian et al.). When also considering the non-LTE results for Fe\,{\sc i}
by Thevenin \& Idiart (1999), the slopes decrease by about 0.06
(King 2000; Israelian et al. 2001), but cf. discussion in 
Sect. \ref{s:fenlte}.}, in stark 
conflict with the [O\,{\sc i}] results.
If confirmed, these results would have far-reaching consequences, as
outlined above. 
The situation is complicated, however, by recent studies of the OH
vibrational-rotational lines in the infrared (IR), which suggest
a nearly flat [O/Fe] in agreement with the [O\,{\sc i}] findings
(Balachandran et al. 2001; Melendez et al. 2001). 

As all previous investigations have been based on 1D hydrostatic model
atmospheres, one may worry about possible systematic errors introduced
by the inherent assumptions of the analyses. Recently the first
3D hydrodynamical model atmospheres of metal-poor stars have been 
constructed (Asplund et al. 1999a), which have very different temperature
structures compared with classical
1D model atmospheres. 
As a consequence, analyses of
temperature sensitive spectral features 
can be suspected to be 
systematically in error if relying on 1D model atmospheres. 
In particular, Asplund et al. (1999a) cautioned that oxygen abundances
of metal-poor stars
derived from 1D studies of OH lines may be strongly overestimated
due to the extreme temperature sensitivity of molecule formation. 
The aim of the present paper is to quantify this suspicion in terms of
O abundances and investigate other possible systematic errors
which may hamper the 3D analysis of OH lines. 
Preliminary results have been presented in Asplund (2001).

\section{3D hydrodynamical model atmospheres}

Realistic {\em ab-initio} 3D, time-dependent simulations of stellar surface
convection form the foundation
for the present study. The same incompressible radiative hydrodynamical code
which previously has been applied successfully to studies of solar
(e.g. Stein \& Nordlund 1998; Asplund et al. 2000a,b) and stellar granulation 
(e.g. Asplund et al. 1999a, Allende Prieto et al. 2001;
Asplund et al., in preparation) has here been
used to construct sequences of 3D model atmospheres 
with varying stellar parameters. 
The equations of mass, momentum and energy conservation together with the
simultaneous treatment of the 3D radiative transfer equation have 
been solved on a Eulerian mesh with 100\,x\,100\,x\,82 gridpoints. 
The physical dimensions of the numerical grid were sufficiently large
to cover many ($\ga 10$) granules simultaneously.
The depth scales have been optimized
to provide the best resolution where it is most needed, i.e. 
in those layers with the steepest
gradients in terms of d$T$/dz and d$^2T$/dz$^2$, which for the solar-type stars
occurs around the visible surface.

Special care has been exercised to include the most appropriate input physics.
In particular, state-of-the-art equation-of-state (Mihalas et al. 1988),
which includes the effects of ionization, excitation and dissociation
of the most important atoms and molecules, and relevant continuous
(Gustafsson et al. 1975 with subsequent updates) and line (Kurucz 1993)
opacities have been employed. 
During the convection simulations, the 3D radiative transfer
is solved for in total eight inclined rays under the simplifying assumptions
of LTE ($S_\nu = B_\nu$) 
and grouping of the opacities into four bins (Nordlund 1982).
At regular intervals during the simulations, 
the accuracy of the opacity binning technique is verified 
by solving the full monochromatic radiative transfer (about 2700 wavelength
points) in the 1.5D approximation, i.e. treating each vertical column as
a separate 1D model atmosphere and ignoring all horizontal radiative 
transfer effects. 
Further details on the numerical procedures of the simulations 
may be found in Stein \& Nordlund (1998). 

\begin{table}[t]
\caption{Details of the 3D hydrodynamical model atmospheres}
\label{t:simulations}
\begin{tabular}{lcccc} 
 \hline 
\noalign{\smallskip}
$<T_{\rm eff}>^{\rm a}$ & log\,$g$ & [Fe/H] & x,y,z-dimensions 
& time$^{\rm b}$ \\
 $$[K] & [cgs] & & [Mm] & [min] 
\smallskip \\
\hline 
\noalign{\smallskip}
$5767\pm21$ & 4.44 & $+0.0$ & 6.0\,x\,6.0\,x\,3.7 & 50 \\
$5822\pm11$ & 4.44 & $-1.0$ & 6.0\,x\,6.0\,x\,3.7 & 50 \\
$5837\pm8$ & 4.44 & $-2.0$ & 6.0\,x\,6.0\,x\,3.7 & 40 \\
$5890\pm8$ & 4.44 & $-3.0$ & 6.0\,x\,6.0\,x\,3.7 & 40 \\
$6191\pm37$ & 4.04 & $+0.0$ & 21.4\,x\,21.4\,x\,8.7 & 60 \\
$6180\pm20$ & 4.04 & $-1.0$ & 21.4\,x\,21.4\,x\,8.7 & 60 \\
$6178\pm17$ & 4.04 & $-2.0$ & 21.4\,x\,21.4\,x\,8.7 & 60 \\
$6205\pm15$ & 4.04 & $-3.0$ & 21.4\,x\,21.4\,x\,8.7 & 60 \\
\hline
\end{tabular}
\begin{list}{}{}
\item[$^{\rm a}$] The temporal average and standard deviation 
of the emergent $T_{\rm eff}$ of the different convection simulations
\item[$^{\rm b}$] The time coverage of the part of the simulation used for the 
spectral line calculations, while the full convection simulations are 
much longer extending over several convective turn-over time-scales
\end{list}
\end{table}

For the present purpose, two sequences of 3D model atmospheres
have been constructed. The first series of models (here: the solar sequence)
correspond to the Sun ($T_{\rm eff} \simeq 5800$\,K
\footnote{Since the entropy of the inflowing gas at the lower boundary
has replaced $T_{\rm eff}$ as an independent input parameter
in 3D convection simulations, the resulting $T_{\rm eff}$ 
varies slightly in time due to the evolution of individual granules.
In order to obtain a specific temporally averaged
value for the emergent $T_{\rm eff}$,
a careful and very time-consuming fine-tuning of the inflowing entropy
would be required. Since we are only interested in a differential
comparison between 3D and 1D, we have not attempted to obtain exactly
the solar $T_{\rm eff}$ for the individual simulations and instead
settled for values in reasonable proximity of the targeted $T_{\rm eff}$,
cf. Table \ref{t:simulations}. Naturally, the comparison 1D model
atmospheres have the same $T_{\rm eff}$ as the final 3D average
to isolate the granulation effects from differences in input parameters.}, 
log\,$g = 4.44$\,[cgs])
but with a range of metallicities ([Fe/H]\,$=0.0$, $-1.0$, $-2.0$ and $-3.0$),
while the second suite (here: the turn-off sequence)
correspond to typical turn-off stars
($T_{\rm eff} \simeq 6200$\,K, log\,$g = 4.04$\,[cgs]), again with varying
metal content ([Fe/H]\,$=0.0$, $-1.0$, $-2.0$ and $-3.0$).
Some details of the simulations are given in Table \ref{t:simulations}. 
The individual elemental abundances have been taken from 
Grevesse \& Sauval (1998) scaled appropriately to the relevant [Fe/H]
with no $\alpha$-element enhancements. 
As long as the comparison is strictly differential between the 3D and
1D model atmospheres as in our case the omission of 
$\alpha$-element enhancements will be negligible for the spectral line
formation. 
The initial snapshots for the simulations were taken from simulations
of lower numerical resolution (50\,x\,50\,x\,82), which had been run
sufficiently long times to allow 
thermal relaxation and a statistically steady state at the
wanted $T_{\rm eff}$ to be established. 
The initial snapshots for these 
lower resolution runs in turn were obtained from a previous solar
simulation (Stein \& Nordlund 1998) scaled appropriately to the new
stellar parameters using the experience from 1D hydrostatic stellar models
and the entropy variations in 2D hydrodynamical model atmospheres
(Ludwig et al. 1999; Freytag et al. 1999).
Classical 1D, hydrostatic {\sc marcs} model atmospheres (Asplund et al. 1997)
with identical input parameters and chemical compositions as the 3D simulations
have been constructed to allow a differential comparison in terms of
spectral line formation. Whether 3D models indeed should be assigned the same
$T_{\rm eff}$ as a corresponding 1D model atmosphere is investigated
in Sect. \ref{s:teff} when comparing the predictions for the infrared
flux method (IRFM) with the two types of models.  

Although the details of the convection properties of the present and other
simulations for late-type stars
will be described elsewhere (Asplund et al., in preparation),
the most important metallicity effects on the resulting photospheric
structures are briefly discussed here in order to understand the 
impact of the 3D models on the spectral line formation of the OH lines.
While the temperature remains close to the radiative
equilibrium value at solar metallicities and 
mild metal-deficiencies ([Fe/H]$\ga -1.0$),
the temperature in the outer layers depart significantly from it at 
lower metallicities
(Asplund et al. 1999a). 
The temperature
in this optically thin region is determined from a competition between
adiabatic cooling and radiative heating. The latter mechanism arises 
when continuum photons released at deeper layers are reabsorbed in 
spectral lines. At solar metallicities the abundant spectral lines
succeed in providing 
sufficient radiative heating to balance the adiabatic cooling and
keeping the average temperature close to the radiative equilibrium
expectation. At progressively lower metallicities, the 
available spectral lines become fewer and weaker, which allows cooling
to dominate more. As a consequence, balance is not restored unless the
temperature is much below the radiative equilibrium value.
Thus, in hydrodynamical model atmospheres spectral lines have the
opposite effect compared to in hydrostatic model atmospheres which
enforce radiative equilibrium: {\em spectral lines cause surface heating}.
The effects of the lower temperatures are first visible at the
outermost layers but move towards deeper layers at progressively
lower metallicities.

When viewed as a function of optical depth instead of geometrical height,
the low surface temperatures
are less pronounced since the temperature sensitivity of the continuous
opacities partly hides the effect (Asplund et al. 1999a); 
lower temperatures also imply smaller
opacities and optical depths. Nevertheless the low temperatures reach
typical line-forming regions at low metallicities. Furthermore, 
the cool outer layers shift the whole line formation outwards for
temperature sensitive features such as molecular lines. 
For example, at log\,$\tau_{500} = -3$ the average temperature difference 
between 3D and 1D models can reach
1000\,K and thus have a profound effect on the lines 
sensitive to those layers.

Differences in the gas and electron pressures could also 
influence spectral line
formation. With the exception of some additional contribution from 
turbulent pressure in the convective overshoot region around the
visual surface, the total pressure is well approximated by the
gas pressure in the photosphere for late-type dwarfs. 
Since even in the 3D hydrodynamical
models the photosphere is typically not too far from 
hydrostatic equilibrium, the pressure scale height 
$H_{\rm P} = -({\rm d ln} P/{\rm d}r)^{-1} \approx P_{\rm gas}/g\rho \propto T$ 
and thus the resulting 3D pressure structure tend to be lower
than the corresponding 1D value at a given geometric height
at low metallicities.
Naturally, the lower temperatures have an even greater
impact on the electron pressure.
The lower gas and electron pressures in metal-poor 
3D model atmospheres will affect mainly
lines that are considered gravity sensitive.

\section{Spectral line formation in 3D model atmospheres}

The OH line calculations with the 3D hydrodynamical model 
atmospheres follow the same procedure as in other recent investigations
of the influence of granulation on stellar spectroscopy
(Asplund et al. 1999a,b, 2000a,b,c; Asplund 2000a,b, 2001; 
Asplund \& Carlsson 2001; Allende Prieto et al. 2001; 
Nissen et al. 2000, 2001; Primas et al. 2000).
The 3D convection simulations, which extend down to the essentially 
adiabatic layers well below the visible photosphere, 
were interpolated to a finer vertical depth-scale 
although with the same number of depth-points to
improve the numerical accuracy in the spectral synthesis. The average
continuum optical depth range typically beyond log\,$\tau_{500} \ga 2$
and above log\,$\tau_{500} \la -5$, in order to minimize the influence 
from the artificial top and bottom boundaries.
Prior to the line transfer calculations
the horizontal resolution was decreased from  
100\,x\,100\,x\,82 to 50\,x\,50\,x\,82 to ease the computational burden; 
various tests ensured that the procedure did not introduce any differences in
the spatially and temporally averaged line profiles. 
From the full convection simulations, which cover several convective turn-over 
time-scales, representative sequences of typically one hour stellar time 
with snapshots every 30\,s were selected for the spectral syntheses.
In terms of derived oxygen abundances, the number of snapshots were sufficient
to provide statistically significant results 
($\Delta {\rm log} \epsilon_{\rm O} < 0.01$\,dex), 
as verified by test calculations with shorter time sequences.
The mean $T_{\rm eff}$'s for the shorter simulation sequences are given in
Table \ref{t:simulations}. 

The spectral line formation in 3D model atmospheres 
was performed under the assumption of LTE.
Thus, the OH number densities were computed from instantaneous molecular
equilibrium and Saha ionization and Boltzmann excitation balances. 
The line source
function was approximated with the Planck function ($S_\nu = B_\nu$).
The applicability of these strong assumptions is further investigated
in Sects. \ref{s:ce} and \ref{s:nlte}. 
The OH molecular equilibrium was computed with scaled solar abundances
with the exception of the oxygen abundance. The removal of available
oxygen atoms due to CO formation was taken into account but 
the effect was found to be negligible.
Flux profiles were computed
for two typical OH UV lines (313.9 and 316.7\,nm) from solving the
radiative transfer for in total 17 inclined rays (four $\mu$-angles
and four $\varphi$-angles plus the vertical $\mu=1.0$). 
The line transition data for the OH lines
was taken from Israelian et al. (1998); we emphasize that in a
differential 3D-1D comparison such as ours the exact choices of for
example the $gf$-values are not important.
In addition, we have included six of the Fe\,{\sc ii} lines
used by Nissen et al. (2001) in their study of [O/Fe] from the [O\,{\sc i}]
line to quantify the corresponding impact on Fe abundances.
The background continuous opacities were calculated using the
Uppsala opacity package (Gustafsson et al. 1975 with subsequent updates).
Since the Doppler shifts introduced by the convective velocity field 
are fully accounted for, no microturbulence or macroturbulence 
parameters enter the 3D line calculations (Asplund et al. 2000b,c).
It is noteworthy that none of the various free parameters 
hampering 1D analyses (e.g. mixing length parameters, micro- and 
macroturbulence) are necessary with 3D hydrodynamical model atmospheres.
For the 1D calculations, a microturbulence of 
1.0\,km\,s$^{-1}$ 
has been assumed for the solar and turn-off sequences;
the choice of microturbulence, however, is only of some significance for
the solar metallicity models while unimportant at low metallicities.

The main advantage of limiting the comparison to a strictly differential
study of the 3D and 1D predictions is that the uncertainties in 
e.g. the absolute transition probabilities, equivalent widths, blends,
continuum placement, missing UV opacities, 
$T_{\rm eff}$-calibration and the solar oxygen abundance can be avoided. 
As a consequence we are not able to determine accurate {\em absolute}
stellar oxygen abundances.
Instead we only attempt to address the question of whether there are
systematic errors in analyses of the UV OH lines when relying 
on classical 1D model atmospheres. 
A star-by-star analysis of high quality OH observations 
is left for a future paper.

\section{3D LTE oxygen abundances derived from UV OH lines
\label{s:oh}}

Molecule formation is extremely temperature sensitive (as long 
as the molecule in question is a trace element for its atomic constituents,
which is the case for OH in the stars studied here).
For conditions typical of the line-forming regions in the Sun,
the LTE number density of OH, $N_{\rm OH}$, is proportional 
to $\approx T^{-12}$ as estimated from our adopted equation-of-state
and molecular balance routines. 
This high degree of non-linearity makes molecular lines 
very susceptible to errors in the adopted temperature structures
of the model atmospheres. In particular, the very different temperature
structures of typical convective up- and downflows compared with 
classical 1D model atmospheres (i.e. up- and downflows can
{\em not} be represented by two theoretical hydrostatic model atmospheres with 
different $T_{\rm eff}$, since granules have a much steeper
temperature gradient than intergranular lanes), 
can be suspected to significantly
influence the molecular number densities and line formation.

\begin{figure}[t]
\resizebox{\hsize}{!}{\includegraphics{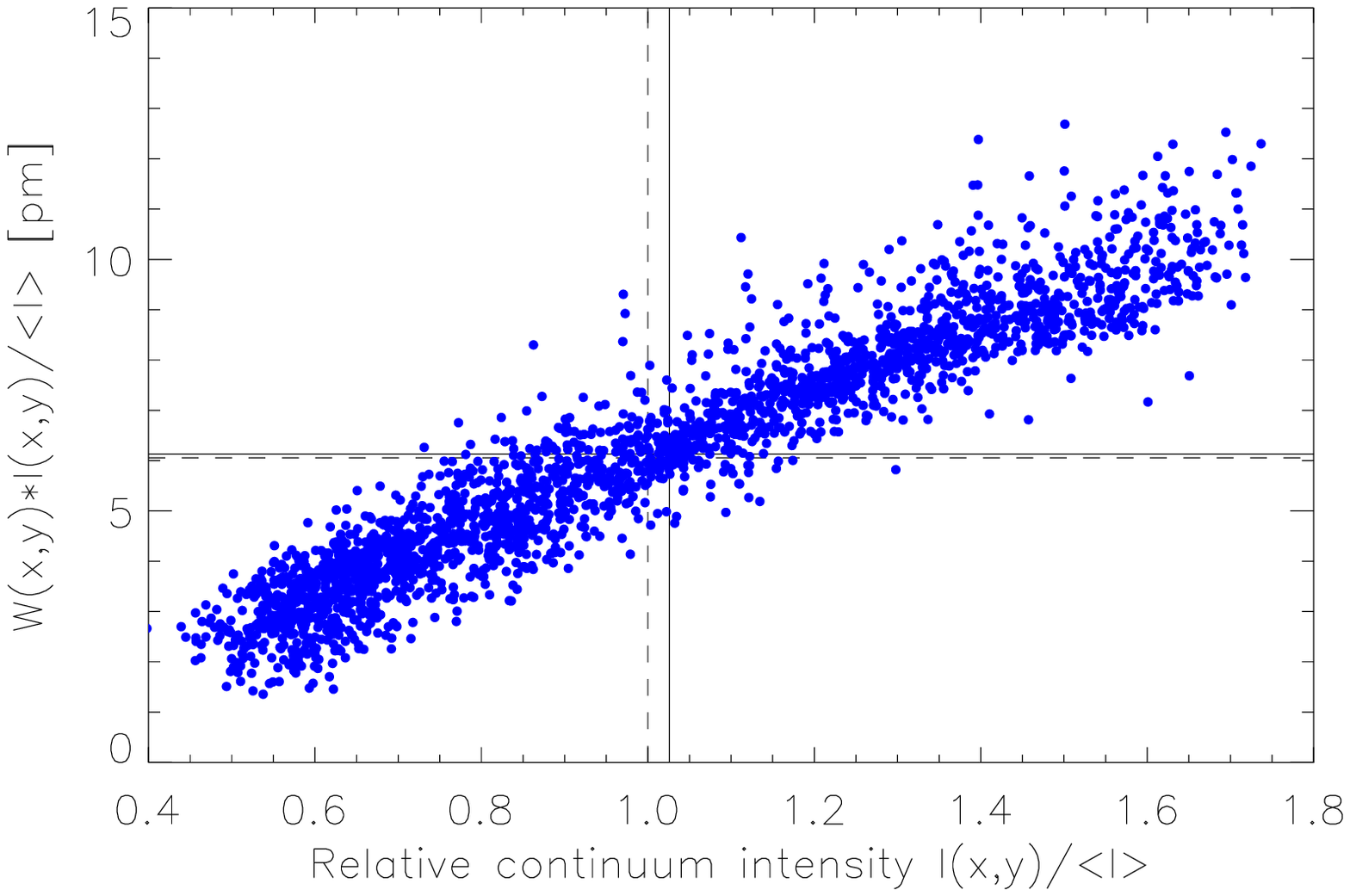}}
\resizebox{\hsize}{!}{\includegraphics{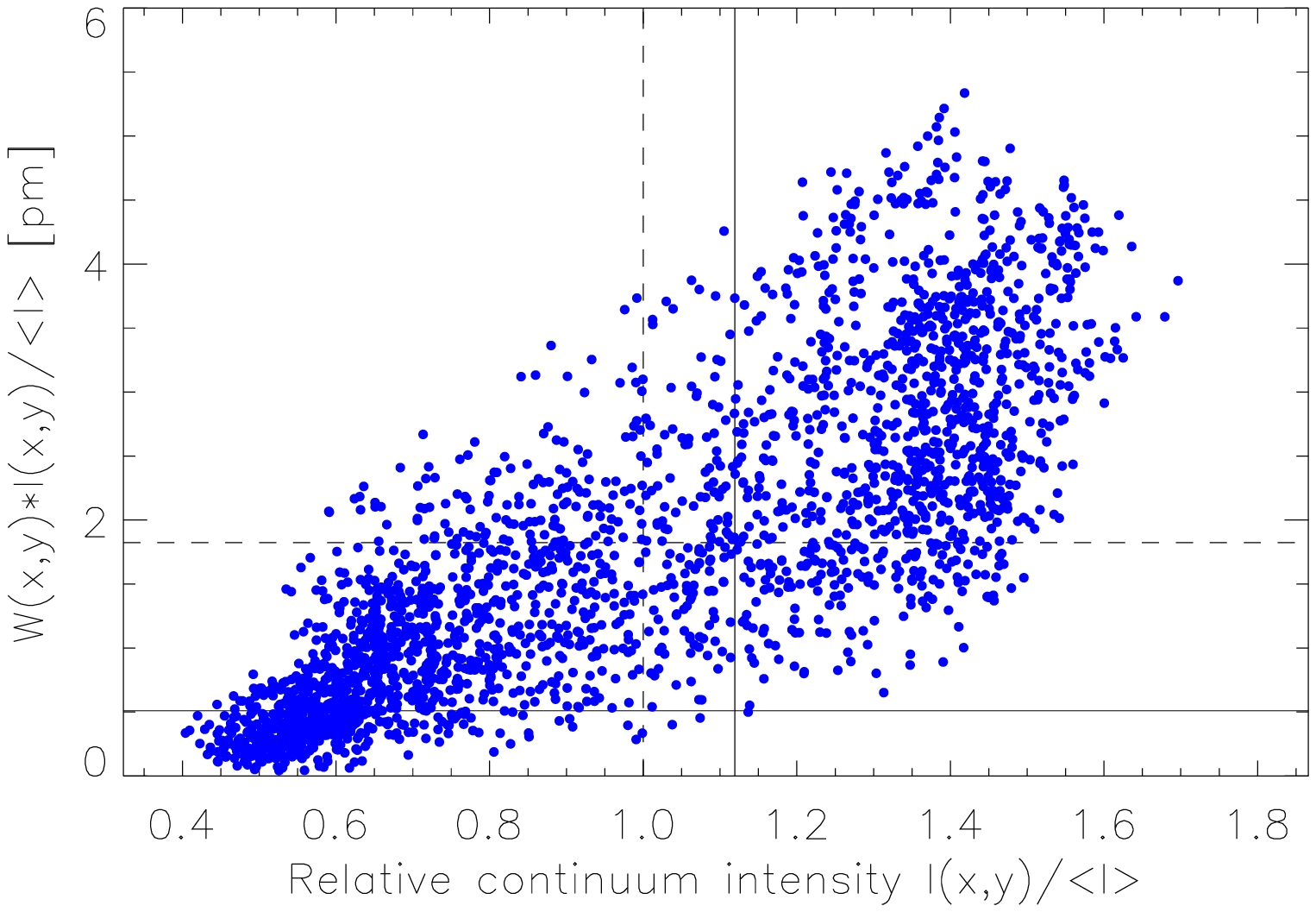}}
\caption{The contribution to the equivalent width, 
$W_\lambda {\rm (x,y)} \cdot 
I^{\rm cont}_\lambda {\rm (x,y)}/\langle I^{\rm cont}\rangle$, 
of the OH\,313.9\,nm line across the stellar surface for the 
[Fe/H]$ = +0.0$ ({\it Upper panel}) and  
[Fe/H]$ = -3.0$ ({\it Lower panel}) models in the solar sequence
as a function of the relative continuum intensity 
$I^{\rm cont}_\lambda{\rm (x,y)}/\langle I^{\rm cont}\rangle$ at
313.9\,nm.
The spatially resolved intensity profiles have been computed for $\mu = 1.0$.
The horizontal and vertical dashed lines denote the mean equivalent widths
and continuum intensities with 3D model atmospheres, 
while the solid lines mark
the corresponding 1D predictions using the same O abundances as in 3D. 
Note that the here shown 3D calculations only refer to one 
snapshot each for the two models and not the whole simulation sequences
}
         \label{f:Wxy}
\end{figure}

Under the assumption of LTE adopted here, $N_{\rm OH}$ depend only
on the instantaneous local temperature. 
In general the temperature contrast reverses
in the convective overshoot region some distance above the continuum
forming layers, i.e. gas above the warm, upflowing granules tend to be
{\em cooler} than average (e.g. Stein \& Nordlund 1998).
In particular
in metal-poor stars the horizontal temperature contrast is very large
due to the weak coupling between the gas and the radiation field
(Asplund et al., in preparation). In general the temperatures at the high
atmospheric layers are very low but occasionally the compression from
converging gas flows or shocks can rise the temperature to the 
radiative equilibrium value or even above it. 
As a consequence, the UV OH LTE line strengths across the stellar granulation
are expected to be stronger in the granules 
but with pronounced scatter 
when viewed nearly face-on, as confirmed by the
3D line calculations (Fig. \ref{f:Wxy}). 
The spatially and temporally averaged OH flux profiles are 
therefore strongly biased towards
upflowing regions due to their higher continuum intensities, 
steeper temperature gradients and larger area coverage
\footnote{Note that the opposite is true for the OH pure rotational lines
in the IR since these lines and their surrounding continuum are formed
at significantly higher layers where the temperature contrast is
reversed relative to the visual continuum 
(cf. Figs. 1 and 7 in Kiselman \& Nordlund 1995)}.

\begin{figure}[t]
\resizebox{\hsize}{!}{\includegraphics{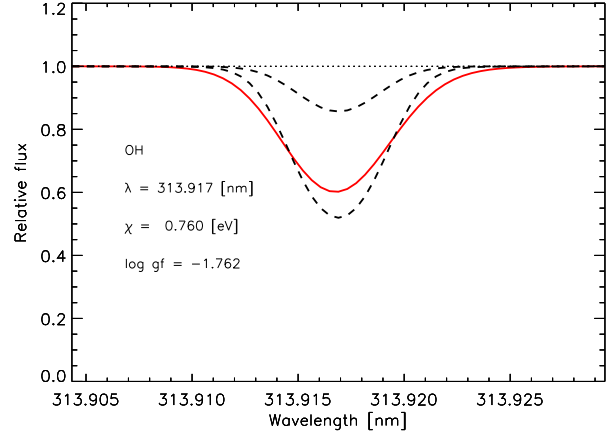}}
\caption{The predicted profile for the OH\,313.9nm line using the
3D hydrodynamical model atmosphere for   
Teff=5800\,K, logg=4.44\,[cgs], [Fe/H]=-3.0 with an adopted 
O abundance of log\,$\epsilon_{\rm O} = 6.4$ (solid line).
In comparison the corresponding 1D profile with the same abundance
is much weaker (upper dashed line). Only with log\,$\epsilon_{\rm O} = 7.05$
(lower dashed line) 
is the 1D profile comparable in strength to the 3D profile.
In no cases have extra broadening in the form of rotation or macroturbulence
been applied}
         \label{f:prof}
\end{figure}

\begin{table}[t]
\caption{Comparison of the oxygen LTE abundances derived with 1D 
hydrostatic and 3D hydrodynamical model atmospheres. The 3D O abundances
are those which reproduce the equivalent widths computed using a 1D model 
atmosphere, a microturbulence of $\xi_{\rm turb} = 1.0$\,km\,s$^{-1}$ 
and the oxygen abundances given in the fourth column
}
\label{t:oh}
\begin{tabular}{lccccc} 
 \hline 
\noalign{\smallskip}
$T_{\rm eff}$ & log\,$g$ & [Fe/H] & log\,$\epsilon_{\rm O, 1D}$ & 
log\,$\epsilon_{\rm O, 3D}$ & 
log\,$\epsilon_{\rm O, 3D}$ \\
 $$[K] & [cgs] & & & 313.9\,nm$^{\rm a}$ & 316.7\,nm$^{\rm b}$ 
\smallskip \\
\hline 
\noalign{\smallskip}
5767 & 4.44 & $+0.0$ & 8.90 & 8.88 & 8.89 \\
5822 & 4.44 & $-1.0$ & 8.30 & 8.03 & 8.07 \\
5837 & 4.44 & $-2.0$ & 7.70 & 7.19 & 7.25 \\
5890 & 4.44 & $-3.0$ & 7.10 & 6.44 & 6.52 \\
6191 & 4.04 & $+0.0$ & 8.90 & 8.80 & 8.82 \\
6180 & 4.04 & $-1.0$ & 8.30 & 7.97 & 8.00 \\
6178 & 4.04 & $-2.0$ & 7.70 & 6.90 & 7.01 \\
6205 & 4.04 & $-3.0$ & 7.10 & 6.08 & 6.22 \\
HM, int.$^{\rm c}$ & 4.44 & $+0.0$ & 8.90 & 8.69 & 8.72 \\
HM, flux$^{\rm c}$ & 4.44 & $+0.0$ & 8.90 & 8.64 & 8.69 \\
\hline
\end{tabular}
\begin{list}{}{}
\item[$^{\rm a}$] log\,$gf = -1.76$, $\chi_{\rm exc} = 0.76$\,eV 
(Israelian et al. 1998)
\item[$^{\rm b}$] log\,$gf = -1.69$, $\chi_{\rm exc} = 1.11$\,eV
(Israelian et al. 1998)
\item[$^{\rm c}$] Using the Holweger-M\"uller (1974) 1D semi-empirical
model atmosphere for the solar case instead of a {\sc marcs} solar model, 
either using intensity or flux line profiles
\end{list}
\end{table}

Typical resulting 
3D and 1D line profiles are shown in 
Fig. \ref{f:prof}, which clearly illustrates the influence of 3D models
on the derived O abundances from UV OH lines at low metallicities. 
The much lower temperatures and larger concentration of OH molecules
in the 3D model atmospheres of metal-poor stars result in far
stronger OH lines than with 1D model atmospheres while the difference
is much less accentuated at solar metallicities, in accordance with
the behaviour of the temperature structures.
Table \ref{t:oh} lists the O abundances of the two
OH lines for the two types of model atmospheres.
The 3D O abundances are those which reproduce the equivalent widths
computed using 1D model atmospheres and an adopted 
[O/Fe]\,$= -0.40 \cdot $[Fe/H] trend
(Israelian et al. 1998, 2001; Boesgaard et al. 1999). 
A less steep [O/Fe] trend would
have resulted in slightly smaller granulation effects due to
the shifting of the line-formation region inwards where the low 
temperatures in the 3D models are less pronounced.
Similarly, the impact of the 3D models is greater for the OH 313.9\,nm
line than for the OH 316.7\,nm line in accordance with the lower
excitation potential and larger line strength of the former.
The in general larger granulation 
corrections for the hotter models are in agreement with their larger 
temperature differences and the high non-linearity of the OH line
formation. 

For completeness we have carried out an identical 3D-1D comparison
for six Fe\,{\sc ii} lines. The resulting granulation corrections
are listed in Table \ref{t:fe}. As expected (Asplund et al. 1999a),
the impact of 3D model atmospheres are relatively minor on the
Fe\,{\sc ii} lines since they are formed in deep atmospheric
layers and therefore not sensitive to the low temperatures 
encountered in the upper layers in the 3D models. In terms
of Fe abundance, the difference between the 1D and 3D predictions
amounts to $\la 0.1$\,dex.

\begin{figure}[t]
\resizebox{\hsize}{!}{\includegraphics{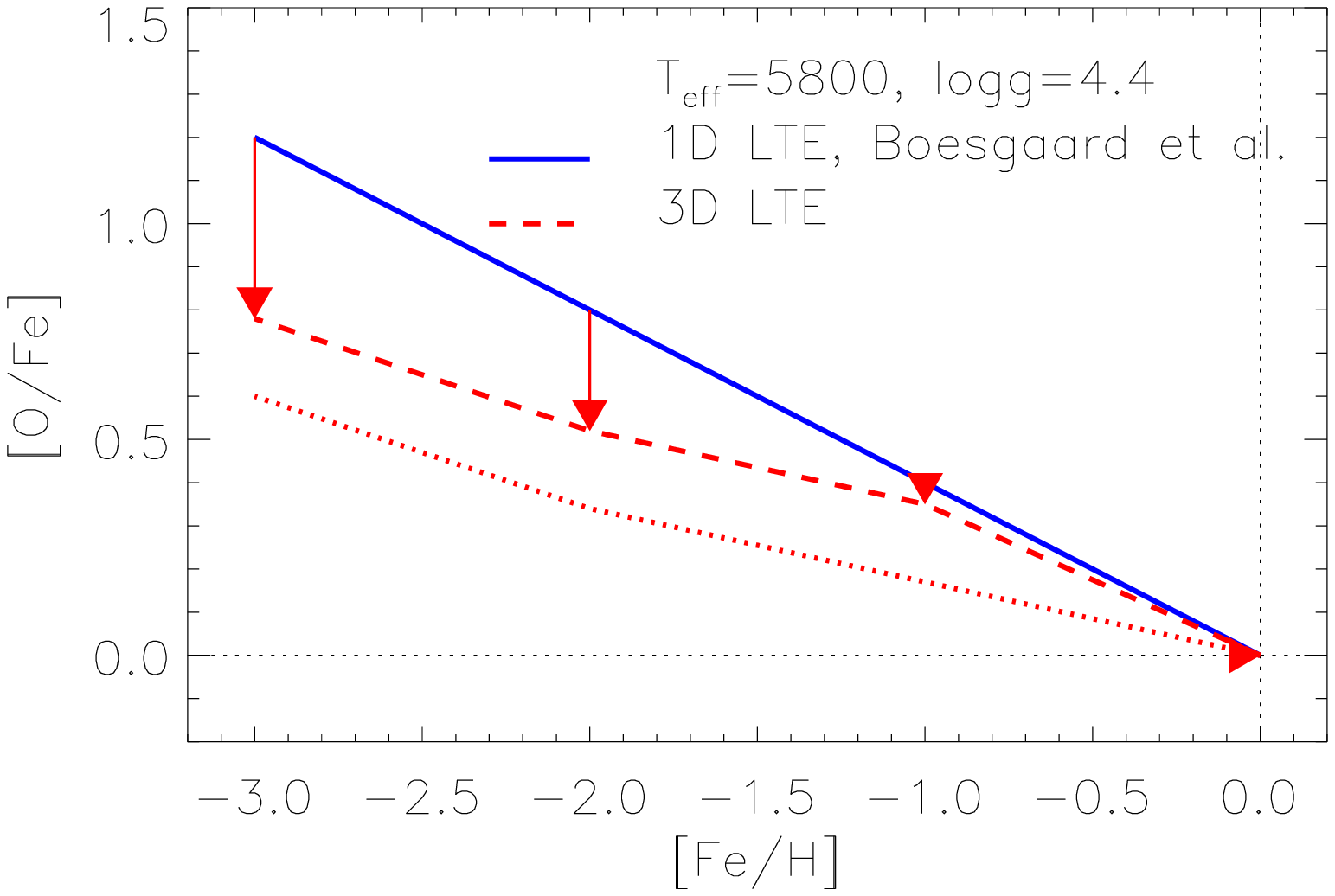}}
\resizebox{\hsize}{!}{\includegraphics{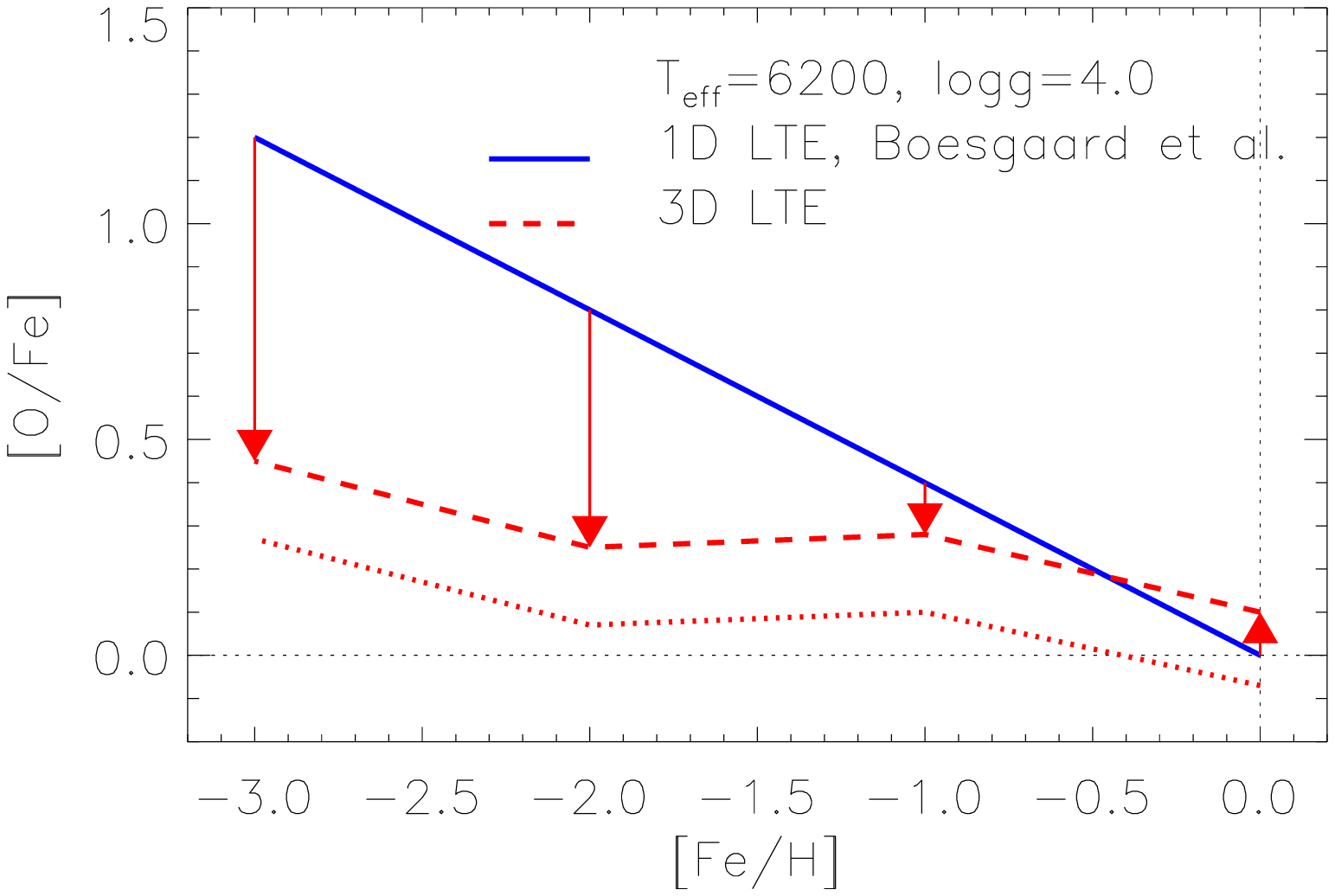}}
\caption{The impact of 3D model atmospheres on the [O/Fe] trend with
metallicity for the solar ({\em Upper panel}) and turn-off ({\em Lower
panel}) sequences. 
The solid line represents the 1D LTE result of Boesgaard et al.
(1999) based on only the UV OH lines and the
King (1993) $T_{\rm eff}$-scale: [O/Fe]\,$= -0.40 \cdot $[Fe/H].
The dashed and dotted lines denote the typical O abundance corrections 
in a 3D LTE analysis (average for the OH 313.9 and 316.7\,nm lines)
compared with the 1D case depending on the 
choice for the solar calibration. For the former the 
Holweger-M\"uller (1974) semi-empirical solar atmosphere
with intensity profiles has been used while
for the latter a theoretical {\sc marcs} model atmosphere together
with flux profiles have been adopted; for all other 1D models theoretical
{\sc marcs} models have been utilised. Hence, in the upper panel the
granulation correction for the [Fe/H]=0.0 model is forced to disappear
since it corresponds to the Sun
}
         \label{f:OFe}
\end{figure}

\begin{figure}[t]
\resizebox{\hsize}{!}{\includegraphics{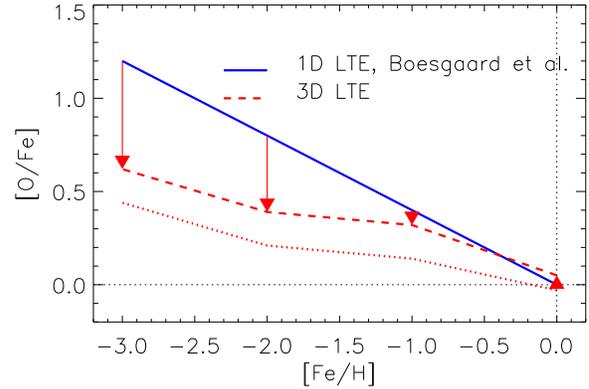}}
\caption{The impact of the 3D LTE analysis of the OH lines on the
[O/Fe] trend in metal-poor stars when averaging the granulation 
corrections for the solar and turn-off sequences for the two
OH lines 313.9 and 316.7\,nm. As in Fig. \ref{f:OFe}, both the cases 
when using the Holweger-M\"uller (1974) model atmosphere (dashed line)
and the {\sc marcs} model atmosphere (dotted line) for the solar
calibration are shown. It should be emphasized that 
the assumption of LTE in the molecule formation and radiative transfer may
skew the results 
according to the discussion in Sect. \ref{s:errors}
}
         \label{f:OFe_average}
\end{figure}

The strong metallicity dependence of the abundance corrections
has a profound impact on the use of UV OH lines as
O diagnostic in metal-poor stars and 
cast serious doubts on 
recent claimed linear trends in [O/Fe] towards lower metallicities
based on 1D LTE analyses 
(Israelian et al. 1998, 2001; Boesgaard et al. 1999).
In Fig. \ref{f:OFe} the average granulation corrections of the two
investigated OH lines for the solar and turn-off sequences are shown,
while Fig. \ref{f:OFe_average} presents the final mean corrections of
the different $T_{\rm eff}$ simulations
\footnote{The results shown in Fig. \ref{f:OFe_average} 
differ slightly from those
presented in Asplund (2001) due to the inclusion of additional
lines and simulations and use of more temporally
extended 3D model atmospheres.}.
The correction of the 1D LTE result due to the use of 3D model
atmospheres is here defined as relative to the solar calibration, i.e.
the abundance differences in Table \ref{t:oh} are subtracted with
the corresponding effect for the Sun since [O/Fe] ratios are studied.
This procedure is similar to the 
common practice of determining astrophysical $gf$-values from the Sun,
as was also the basis for the works by Nissen et al. (1994),
Israelian et al. (1998, 2001)
and Boesgaard et al. (1999).
For the Sun, two possible calibrations are possible using either the
Holweger-M\"uller (1974) semi-empirical model atmosphere or the
theoretical {\sc marcs} model as the 1D representation of the solar
photosphere, yielding two possible 3D trends in Figs. \ref{f:OFe}
and \ref{f:OFe_average}.  
The difference between the two trends therefore simply reflects the 
difference in derived solar O abundances when relying on the two types
of 1D model atmospheres.
It should be noted that the granulation corrections for Fe\,{\sc ii}
lines presented in Table \ref{t:fe} are {\em not} included
in Figs. \ref{f:OFe} and \ref{f:OFe_average}, since the adopted 
[O/Fe] trend with metallicity originates from analyses of
Fe\,{\sc i} lines (Boesgaard et al. 1999; Israelian et al. 1998, 2001,
cf. discussion in Sect. \ref{s:fenlte}). According to Table \ref{t:fe},
the inclusion of the Fe\,{\sc ii}
results would bring down the 3D LTE [O/Fe] results further by 
$\la 0.1$\,dex at the lowest metallicities.

\begin{table}[t]
\caption{Comparison of the iron LTE abundances derived with 1D 
hydrostatic and 3D hydrodynamical model atmospheres from Fe\,{\sc ii} lines. 
The 3D Fe abundances
are those which reproduce the equivalent widths computed using a 
1D model atmosphere, a microturbulence of $\xi_{\rm turb} = 1.0$\,km\,s$^{-1}$ 
and the Fe abundances given in the fourth column
}
\label{t:fe}
\begin{tabular}{lccccc} 
 \hline 
\noalign{\smallskip}
$T_{\rm eff}$ & log\,$g$ & [Fe/H] & log\,$\epsilon_{\rm Fe, 1D}^{\rm a}$ & 
log\,$\epsilon_{\rm Fe, 3D}^{\rm a}$ & $\Delta {\rm log} \epsilon_{\rm Fe}$ \\
 $$[K] & [cgs] & & &  
\smallskip \\
\hline 
\noalign{\smallskip}
5767 & 4.44 & $+0.0$ & 7.50 & 7.48 & $-0.02$ \\
5822 & 4.44 & $-1.0$ & 6.50 & 6.54 & $+0.04$ \\
5837 & 4.44 & $-2.0$ & 5.50 & 5.60 & $+0.10$ \\
5890 & 4.44 & $-3.0$ & 4.50 & 4.59 & $+0.09$ \\
6191 & 4.04 & $+0.0$ & 7.50 & 7.49 & $-0.01$ \\
6180 & 4.04 & $-1.0$ & 6.50 & 6.52 & $+0.02$ \\
6178 & 4.04 & $-2.0$ & 5.50 & 5.56 & $+0.06$ \\
6205 & 4.04 & $-3.0$ & 4.50 & 4.57 & $+0.07$ \\
HM, int.$^{\rm b}$ & 4.44 & $+0.0$ & 7.50 & 7.52 & $+0.02$ \\
HM, flux$^{\rm b}$ & 4.44 & $+0.0$ & 7.50 & 7.48 & $-0.02$ \\
\hline
\end{tabular}
\begin{list}{}{}
\item[$^{\rm a}$] The impact of 3D model atmospheres on the derived
stellar metallicities have been investigated for in total 
six Fe\,{\sc ii} lines (614.9, 623.8, 624.7, 641.7, 643.2 and 645.6\,nm). 
In all cases the different Fe\,{\sc ii} lines give the same granulation
corrections to within 0.03\,dex.
\item[$^{\rm b}$] Using the Holweger-M\"uller (1974) 1D semi-empirical
model atmosphere for the solar case instead of a {\sc marcs} solar model, 
either using intensity or flux line profiles
\end{list}
\end{table}

Taken at face value, with the claimed 1D [O/Fe] results and the here
presented granulation corrections, the emerging trend with metallicity is
in fact roughly consistent with the since long advocated [O/Fe] plateau
from [O\,{\sc i}] lines, as seen in Fig. \ref{f:OFe_average}.
However, we caution that such a conclusion is likely premature, as
there still exist inherent assumptions and approximations in the present
3D analysis, most notably the use of instantaneous molecular equilibrium and
LTE radiative transfer (Sects. \ref{s:ce} and \ref{s:nlte}), besides
of course the fact that no direct comparison with observations
on a star-by-star basis has been made. 
Furthermore, the issues of possible missing UV opacities
(Balachandran \& Bell 1998; Bell et al. 2001) and stellar Fe
abundances (Thevenin \& Idiart 1999; King 2000, cf. Sect. \ref{s:fenlte}) 
must be addressed before
safe conclusions can be drawn from OH lines regarding the [O/Fe] behaviour
in metal-poor stars.
Therefore, we here refrain from claiming accordance between
the UV OH and [O\,{\sc i}] results and instead settle to point
out a possible 
serious systematic error affecting recent 1D analyses of OH lines. 
As a result, the case for a continuous linear increase in [O/Fe]
towards lower metallicities is seemingly much weakened.

Na{\"\i}vely, one could expect that 
all OH molecular lines should be similarly affected by
the low temperatures and high molecular number densities in the
3D model atmospheres of metal-poor stars. In view of this, the low
[O/Fe] values found from OH vibrational-rotational lines in the IR
from cool ($T_{\rm eff} \la 5000$\,K) metal-poor giants and dwarfs 
(Balachandran et al. 2001; Melendez et al. 2001)
appear surprising. 
No 3D hydrodynamical model atmospheres are yet available for 
these lower $T_{\rm eff}$ and/or surface gravities while the IR lines
are unpropitiously weak for the stellar parameters of the 
current ($T_{\rm eff} \ga 5800$\,K) suites of models. 
Hence we are unable to confirm or disprove this apparent discordance.
However, we have verified that OH lines in the IR with similar
line strengths as the UV lines suffer from as severe granulation abundance
corrections for the models listed in Table \ref{t:simulations} using
fake OH lines with strongly enhanced transition probabilities. 
Thus, the spectral locations of the transitions are not a 
solution to this conundrum. 
Whether the differences in stellar parameters may be a possible 
resolution is discussed further in Sect. \ref{s:odiagnostics}.

\section{Possible systematic errors affecting the 3D results}
\label{s:errors}

\subsection{Temperature structure in 3D model atmospheres
\label{s:temp}}

The large granulation effects on the derived O abundances for
OH lines in metal-poor stars
hinge crucially on the existence of the low atmospheric
temperatures encountered in the OH line-forming region with
3D model atmospheres. It is therefore natural to inquire whether 
the resulting temperature structures are indeed 
accurate representations of the real stellar atmospheres.
Even if the here employed 3D hydrodynamical model atmospheres
no longer rely on the same simplifying assumptions 
as in classical 1D model atmospheres, 
uncertainties in the temperature structures may still remain.
A consistency check on the 3D atmospheres is available
from similar 2D radiative-hydrodynamical 
simulations (Ludwig et al. 1999; Freytag et al. 1999),
which also produce the distinct sub-radiative equilibrium temperatures.
This consonance is reassuring but not too surprising given the resemblance 
in the underlying assumptions of the two types of simulations.
 
As explained in Sect. \ref{s:oh}, 
the low atmospheric temperatures are natural consequences of no
longer enforcing radiative equilibrium and instead solving explicitly
the time-dependent energy equation. 
This phenomenon is therefore a real physical effect
which must be present in the photospheres of metal-poor stars.
Nevertheless, the magnitude of the effect
may have been overestimated in the present simulations. 
Asplund et al. (1999a) indeed cautioned that the neglect of Doppler shifts
in the treatment of the strong spectral lines in the construction of
the 3D model atmospheres may lead to the radiative heating being
underestimated. 
Work is currently being undertaken to construct  
such further improved 3D model atmospheres.

An alternative avenue to proceed is to design
observational tests to confront the predictions from 3D model atmospheres
with. An often-used method is a detailed comparison of spectral line
asymmetries and shifts (e.g. Asplund et al. 2000a,b; Allende Prieto et al. 2001).
Unfortunately, a similar study of the asymmetries of the OH lines 
is unlikely to give crucial clues due to 
the inevitable blends in the UV. 
Furthermore, only limited guidance is likely to be obtained from studies
of line asymmetries of atomic species since they do not probe the 
same high atmospheric layers as the molecular transitions.

A possible venture to explore is the individual O abundances obtained
from a large sample of OH lines of different excitation potential. 
Since the exact 3D abundance corrections depend
on the transition properties with low-excitation and strong lines being more
influenced by the 3D model atmospheres (Table \ref{t:oh}), different
results are expected compared with 1D model atmospheres. 
An examination of Table 3 in Israelian et al. (1998) indeed reveals that 
low excitation OH lines 
(e.g. 312.39, 312.81, 312.83\,nm, $\chi_{\rm exc} \simeq 0.2$\,eV)
appear to systematically give $\simeq 0.1$\,dex higher abundance than 
the high excitation lines 
(e.g. 316.71, 320.39, 325.55\,nm, $\chi_{\rm exc} \simeq 0.8-1.3$\,eV)
in a 1D analysis, hinting to a possible problem in the 1D temperature
structures at low metallicities {\em relative to the Sun}.
However there is no clear trend with metallicity in the 
differences in the derived
O abundances but the scatter is unfortunately very large.
A proper investigation will therefore require many more lines for a 
larger stellar sample. Furthermore, a prerequisite
is an accurate determination of the stellar parameters, which, in view
of the current disagreement on the $T_{\rm eff}$-scale of metal-poor stars, 
is unlikely to be settled for good in the near future.

\subsection{Effective temperature 
\label{s:teff}}

In order to obtain accurate absolute abundances, not only a realistic
model of the stellar photosphere and a proper understanding of the 
line formation
process are necessary but also appropriate fundamental stellar parameters.
For OH lines, the effective temperature is of special importance.
In Sect. \ref{s:oh} the comparison between the 3D and 1D predictions
were carried out assuming that the relevant $T_{\rm eff}$ should be
the same for the two types of models. 

One may suspect that the presence of temperature inhomogeneities in the
continuum-forming layers should make the emergent flux distribution 
different in 3D model atmospheres compared with homogeneous 1D models,
in particular for metal-poor stars which often are characterized by 
``naked granulation'' 
(Nordlund \& Dravins 1990; Asplund et al., in preparation):
the region of maximum horizontal temperature contrast reaches the visible
surface whereas for example for the Sun these layes are
hidden slightly below the photosphere (Stein \& Nordlund 1998).
This difference will be particularly manifested in UV colours (amounting
to $\la 10$\% in continuum fluxes at 314\,nm, cf. Fig. \ref{f:Wxy}) 
while the effect is minimized at IR wavelengths.  
Since IRFM is our preferred choice for $T_{\rm eff}$-calibrations,
we will here only investigate the impact of 3D models on this method.

The IRFM is designed to compare the observed ratio of 
the total bolometric flux of a star and a  
monochromatic continuum flux at IR wavelengths 
with the corresponding theoretical ratio from model atmospheres
(Blackwell \& Shallis 1977),
In practice, JHK photometry
normally replaces the monochromatic continuum
flux for observational convenience.
Here we follow the original idea by computing 
the spatially and temporally averaged
continuum flux at 2.2\,$\mu$m for the various 3D simulations and compare these
predictions with corresponding ones for 1D model atmospheres with 
$T_{\rm eff, 3D}$ and  $T_{\rm eff, 3D}\pm100$\,K to quantify typical
corrections to $T_{\rm eff}$-estimates based on the IRFM and
classical 1D model atmospheres.
It should be noted that our procedure thus neglects the model atmosphere
dependence in estimating the stellar fluxes {\em outside} the observed
photometric bands to obtain bolometric fluxes 
(Alonso et al. 1995). However, only a small
fraction ($\la 10\%$) of the total flux is carried at those wavelengths for the
F-K stars of interest here and the slight inconsistency of relying on
1D models for this purpose should have a marginal effect on 
the final $T_{\rm eff}$-calibration.

\begin{figure}[t]
\resizebox{\hsize}{!}{\includegraphics{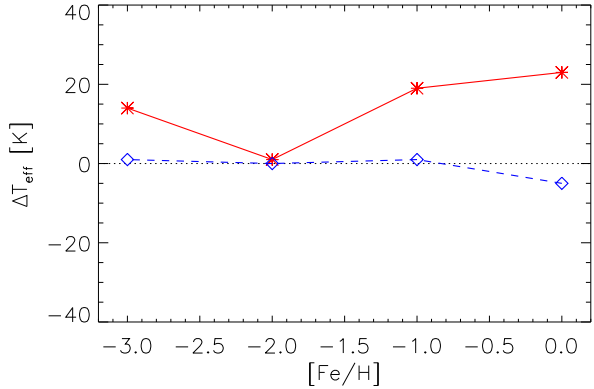}}
\caption{The differences in $T_{\rm eff}$ derived from IRFM  
for the solar (solid line) and turn-off (dashed line) 
sequences of models in the sense $T_{\rm eff, 3D}$- $T_{\rm eff, 1D}$.
The dotted line represent equality between the $T_{\rm eff}$ 
determinations}
         \label{f:deltateff}
\end{figure}

The differences in $T_{\rm eff}$-determinations 
from 3D and 1D model atmospheres are 
shown in Fig. \ref{f:deltateff}. Clearly due to the small model sensitivity 
of the method, the use of 1D models in available IRFM determinations 
does not significantly encumber the results. For the solar sequence the
$T_{\rm eff}$ differences amount to typically $20$\,K while the modifications
for the turn-off sequence are completely negligible.
The smaller effect for the hotter models is in accordance with the
relatively smaller fraction of the total flux emitted at IR wavelengths
and thus smaller sensitivity of the temperature structure of the
adopted model atmospheres.
In view of the
typical observational uncertainties of $\approx 100$\,K currently
attached to IRFM, the errors in IRFM $T_{\rm eff}$ estimates are
still very much dominated by the accuracy of the 
observations rather than the adopted model atmospheres. 

The small influence on $T_{\rm eff}$ translates to only a minor impact 
on the estimated O abundances.
An increase in $T_{\rm eff}$ by $+100$\,K typically implies an
increase of the derived O by 0.2\,dex  
for the UV OH lines (Nissen et al. 1994). Thus, a
new $T_{\rm eff}$-calibration based on 3D model atmospheres is only
expected to introduce a $\la 0.05$\,dex alteration of the 
inferred O abundances.
Furthermore, there is no significant metallicity dependence in the
$T_{\rm eff}$-corrections, which could bias any deduced [O/Fe] trends.

We conclude that 
[O/Fe] determinations will
{\em not} be substantially modified due to changes in the stellar
parameters entailed by the adoption of 3D model atmospheres.

\subsection{Surface gravity
\label{s:logg}}

In addition to a specification of $T_{\rm eff}$ (Sect. \ref{s:teff}),
knowledge of the stellar surface gravity, log\,$g$, is required for 
the spectral synthesis. In Sect. \ref{s:oh} the 3D and 1D 
calculations were performed with identical log\,$g$, which
may not be appropriate.

The best method to determine stellar surface gravities
is to make use of the accurate parallaxes now available from
the Hipparcos mission. By manipulating the familiar relationships
$g \propto M/R^2$ and $L \propto R^2 T_{\rm eff}^4$, the
trigonometric gravities are: 

\begin{equation}
{\rm log} g = 4{\rm log}T_{\rm eff} + {\rm log}M/M_{\sun} +
2{\rm log}\pi + 0.4(V + BC) - 10.51
\end{equation}

\noindent where $M$ is the stellar mass, 
$V$ is the apparent visual magnitude of the star
corrected for extinction, $BC$ is the bolometric correction and
$\pi$ is the parallax (e.g. Nissen et al. 1997).
The additional uncertainty introduced by the use of 3D model
atmospheres instead of classical 1D models enters through the
small systematic differences in $T_{\rm eff}$ and $BC$, while the indirect
effect on the derived mass through isochrone-fitting is negligible.
As seen in Sect. \ref{s:teff}, the difference in $T_{\rm eff}$-calibration
between the two types of models amount to $\pm 20$\,K in the
relevant temperature range, which corresponds to an error of merely
$\pm 0.01$\,dex for log\,$g$. No bolometric corrections have yet been computed
with 3D model atmospheres but considering that $\Delta T_{\rm eff} = 500$\,K
corresponds to $\Delta BC \simeq 0.05$ for $T_{\rm eff} \simeq 6000$\,K
with 1D model atmospheres (e.g. Alonso et al. 1995), 
a very conservative limit to the differential effect of 3D models 
is $\Delta BC \le 0.1$. 
Thus, log\,$g$ determinations using 3D model atmospheres are only
expected to be $\le 0.05$\,dex different from a corresponding 1D analysis
when relying on the parallax-method. We emphasize that this is not
the full uncertainty with this method, which is dominated by the 
errors in the parallax measurements and the $T_{\rm eff}$-calibration,
but only the additional uncertainty relative to the 1D case  
when relying on 3D model atmospheres. In terms of derived O abundances,
$\Delta {\rm log} g \le \pm 0.05$ corresponds to $\Delta (O/H) \le \mp 0.02$\,dex
for the UV OH lines (Nissen et al. 1994).
We can therefore safely conclude that our assumption of identical adopted
surface gravities in the 3D-1D comparison in Sect. \ref{s:oh} will not
impede the conclusions presented therein.

Alternative methods to derive stellar surface gravities are 
available from spectroscopy, in particular by enforcing ionization equilibrium 
or using the pressure-damped wings of strong lines.
Unfortunately they
suffer from several drawbacks which make them less attractive
in analyses with 3D model atmospheres.
In LTE the 3D abundances
derived from Fe\,{\sc i} lines are much lower than in classical
1D analyses (Asplund et al. 1999a), which, if correct, would lead to large
modifications of the derived log\,$g$ from ionization balance.
However, the Fe\,{\sc i} lines are almost certainly seriously
affected by departures from LTE in 3D models, and thus a 3D non-LTE
study of Fe line formation would be required, a very challenging task for
the future.
Indeed, even with 1D model atmospheres departures from LTE make the
ionization gravities discrepant from trigonometric gravities 
(Allende Prieto et al. 1999b, cf. Sect. \ref{s:fenlte}).  
Similarly, the strong lines normally utilised for gravity-determinations
are from species which can be expected to be affected by departures
from LTE in 3D model atmospheres (Mg\,{\sc i}, Ca\,{\sc i}, Fe\,{\sc i}),
again necessitating 3D non-LTE investigations.
Furthermore, at very low metallicities ([Fe/H]~$\la -2$) also the 
strongest lines become 
too weak to accurately probe the photospheric pressure structure
(Fuhrmann 1998). 

\subsection{Metallicity
\label{s:fenlte}}

Stellar metallicities enter into abundance analyses both indirectly 
through their influence on the photospheric structure 
and directly through the use of Fe
as a reference element for abundance ratios. While the dependence on
the former is relatively weak (an error in [Fe/H] as large as
0.4\,dex only implies an uncertainty in the derived O abundance from
OH lines of 0.05\,dex, Nissen et al. 1994), the latter is  
as important as deriving accurate O abundances 
when attempting to trace the evolution of [O/Fe]. 
This obvious fact is, however, often over-looked
with values simply taken from the literature or estimated only from
Fe\,{\sc i} lines with no consideration for departures from LTE.
Since the recent analyses of UV OH lines by Israelian et al. (1998)
and Boesgaard et al. (1999) have utilised Fe\,{\sc i} lines,
their derived [O/Fe] trends may be systematically overestimated
(King 2000; Israelian et al. 2001), which
in turn could influence the 3D-1D comparison presented in Sect. \ref{s:oh}. 
Recently two investigations of non-LTE effects in Fe\,{\sc i} line formation
in metal-poor stars have been published 
(Gratton et al. 1999; Thevenin \& Idiart 1999)
although with discomfortingly discordant results, which deserves 
further scrutiny.  

We tend to view the calculations of Gratton et al. (1999) with some
balanced scepticism. Their incomplete treatment of the high-excitation 
levels, neglect of available quantum mechanical calculations
for the photo-ionization cross-sections 
(from e.g. the {\sc iron} Project, Bautista 1997) and their exceedingly
large cross-sections for inelastic collisions with H, all combine to
ensure a result close to the LTE prediction. Thevenin \& Idiart (1999)
on the other hand adopt more realistic atomic input data but still
suffer from the incomplete handling of the line-blanketing.
Since the main non-LTE effect, over-ionization, feeds
on the UV radiation field it is paramount to address the UV line-blocking
in the calculations of the photo-ionization rates to avoid predicting
too large departures from LTE for Fe\,{\sc i}.
Furthermore, improved quantum mechanical calculations for
the H-collisions are urgently needed to replace the
questionable classical recipe of Drawin (1968).

Until improved non-LTE calculations are available, we
urge that the determinations of stellar Fe abundances to be based on 
Fe\,{\sc ii} lines.
Fe\,{\sc ii} lines are essentially immune to departures from LTE
(e.g. Shchukina \& Trujillo Bueno 2001) 
and as clear from Table \ref{t:fe} are not particularly
affected by the temperature inhomogeneities and different temperature
structures in 3D model atmospheres.  
At this stage, however, we can not exclude that the estimated
[O/Fe] trend with 3D models (Fig. \ref{f:OFe_average}) may even need further 
downward adjustment due to departures from LTE for Fe\,{\sc i} lines,
on which the existing 1D [Fe/H] estimates are based 
(Israelian et al. 1998, 2001; Boesgaard et al. 1999).
Naturally, investigations of departures from LTE for Fe in metal-poor
stars should also be based on the new generation of 
3D model atmospheres.

\subsection{Molecular equilibrium for OH formation
\label{s:ce}}

For the LTE line formation calculations presented in Sect. \ref{s:oh},
the assumption of instantaneous molecular equilibrium has been made
in the computation of the total number density of OH molecules at
different times and locations in the 3D model atmospheres. 
According to Fig. \ref{f:Wxy}, the OH line formation is strongly
biased towards the upflow regions, where the gas is very rapidly
cooled from about 10\,000\,K to about 4000\,K in a relatively
thin zone around continuum optical depth unity. 
This transition occurs on a time-scale of merely a few minutes for
the upflowing material in the Sun, 
which could imply that molecular equilibrium is not established. 
Additionally, photodissociation due to the non-local
radiation field from deeper layers may cause further departures from LTE.
As a result, one would expect that LTE may overestimate the OH content
and therefore that the O LTE abundances may be underestimated.

In principle it is a straight-forward exercise to compute the 
OH molecule formation and the resulting OH number densities 
as a function of time in our 3D model atmospheres by solving 
a set of coupled differential equations corresponding to 
a network of chemical pathways.
However, a major obstacle is the apparent lack of rate
coefficients for the relevant reactions and temperatures, both
experimental and theoretical. 
We have scoured various publically available databases such as UMIST
(Le Teuff et al. 2000)
in search of the necessary rate coefficients but 
with little success in locating data for $T \ga 3000$\,K, which
prevents us from performing the non-LTE chemistry calculations. 

Some guidance to the non-LTE behaviour may still be obtained 
from observations in the absence of detailed computations. 
Uitenbroek (2000a,b) has recently concluded from a comparison of
the observed solar CO line intensities and their
temporal variations with 
calculations based on both 1D hydrodynamical chromospheric simulations
(Carlsson \& Stein 1992, 1995, 1997) 
and 3D model atmospheres similar to those utilised here
(Stein \& Nordlund 1998) that the inherent assumption
of instantaneous molecular equilibrium for CO may not be valid in the Sun. 
This may suggest that a similar phenomenon could also occur for OH. 
If so, it would probably be more pronounced in metal-poor
stars in view of their more rapid and dramatic cooling in 
the photosphere. 

We conclude that it can not be excluded that
departures from molecular equilibrium 
may influence the derived O abundances when
using OH lines and that its magnitude could be metallicity dependent.
Naturally this should be examined further once the necessary
data becomes available.

\subsection{Local thermodynamic equilibrium for OH line formation
\label{s:nlte}}

Besides the assumption of chemical equilibrium for the OH molecule
formation, it should be borne in mind that LTE has also been assumed
in the OH line transfer calculations presented in Sect. \ref{s:oh}.
Due to the vast number of relevant levels and transitions, 
detailed non-LTE calculations for the OH radiative transfer is formidable.
In fact, non-LTE effects for molecules in stellar 
atmospheres is largely unexplored even with 1D model atmospheres. 

We are not aware of any non-LTE studies for OH but CO has attracted 
slightly more attention. 
Uitenbroek (2000b) has recently performed detailed non-LTE calculations
for the CO vibrational-rotational lines in the Sun,
which confirm the insightful prediction by
Hinkle \& Lambert (1975) that the lines are
collisionally controlled and thus that LTE is a good approximation.
Hinkle \& Lambert caution on the other hand  
that electronic molecular transitions, like
the UV OH lines, may be radiatively determined.
If one approximates the UV OH line formation with the two-level
approach with complete redistribution (cf. Mihalas 1978), 
the line source function $S_{\rm l}$ will depend on
the mean intensity averaged over the absorption profile 
$\bar{J_\lambda} = \int \phi_\lambda J_\lambda d\lambda$ 
and Planck function $B_\lambda(T)$ as 
$S_{\rm l} = (1-\epsilon) \bar{J_\lambda} + 
\epsilon B_\lambda(T).$
Here $\epsilon$ is a measure of the photon destruction probability
($0 \le \epsilon \le 1$).
In the UV, $\bar{J_\lambda}$ tend to be larger than $B_\lambda(T)$ 
for weak lines and
therefore $S_{\rm l} > B_\lambda(T)$. As a consequence, one would 
expect the OH lines to be weaker with scattering than in LTE, or,
equivalently, that the derived O abundance will be underestimated in LTE. 

A detailed non-LTE calculation for OH including all the vibrational
and rotational levels is unfortunately 
beyond the scope of the present investigation.
But we have nevertheless attempted to estimate the non-LTE corrections to
the derived O abundances from the UV OH lines using a two-level OH molecule.
Although no doubt unrealistically simplistic to allow accurate quantitative 
estimates of the non-LTE effects, the approach is still expected to yield
qualitatively correct results of the non-LTE behaviour for different
stellar parameters. For the purpose, version 2.2 of 
the statistical equilibrium code {\sc multi} (Carlsson 1986) 
has been used after some minor modifications to
allow treatment of OH molecular lines. The two levels correspond to the
OH 313.9\,nm transition with the same adopted parameters as for the
LTE calculations presented in Sect. \ref{s:oh}. Additionally,
cross-sections for collisions with electrons and hydrogen must be specified.
For the former the classical recipe of van Regemorter (1962) was adopted
in the absence of more appropriate treatments.
For H-collisions the calculations were performed with 
the formula by Drawin (1968) multiplied by a variable factor $x$.
It should be noted that the Drawin formula was developed for atoms
and it is not clear whether it is at all applicable for molecules.
Although often used in non-LTE calculations for late-type stars
(cf. discussion in Kiselman 2001),
evidence is now mounting that the simple-minded approach by Drawin 
severely overestimates the importance of H-collisions by at least
three orders of magnitude for atomic transitions
(Fleck et al. 1991;
Belyayev et al. 1999).
The inclusions of vibrational and rotational 
sub-levels and line-blanketing in the UV radiation field 
are also expected to diminish the non-LTE effects due
to stronger collisional quenching and decreased
$\bar{J_\lambda}/B_\lambda(T)$ ratios in the line-forming region. 

\begin{table}[t]
\caption{The 1D non-LTE O abundance corrections when using a two-level
model-molecule to study the UV OH line formation. The last
two 1D models correspond to HD\,140283 and G64-12, respectively.}
\label{t:nlte}
\begin{tabular}{lccccc} 
 \hline 
\noalign{\smallskip}
$T_{\rm eff}$ & log\,$g$ & [Fe/H] & [O/Fe]$^{\rm a}$ & $x^{\rm b}$ & 
$\Delta {\rm log} \epsilon_{\rm O}^{\rm c}$ \smallskip \\
$ $[K] & [cgs] & &  [dex] & & [dex]  \smallskip \\
\hline 
\noalign{\smallskip}
5780 & 4.44 & $+0.0$ &    0.0 & 0.0   & $+0.18$ \\
     &      &        &        & 0.01  & $+0.18$ \\
     &      &        &        & 0.1   & $+0.18$ \\
     &      &        &        & 1.0   & $+0.17$ \\
     &      &        &        & 10    & $+0.12$ \\
     &      &        &        & 100   & $+0.03$ \\
5780 & 4.44 & $-1.0$ & $+0.4$ & 0.0   & $+0.20$ \\
     &      &        &        & 0.01  & $+0.20$ \\
     &      &        &        & 1.0   & $+0.17$ \\
5780 & 4.44 & $-2.0$ & $+0.8$ & 0.0   & $+0.25$ \\
     &      &        &        & 0.01  & $+0.25$ \\
     &      &        &        & 1.0   & $+0.22$ \\
5780 & 4.44 & $-3.0$ & $+1.2$ & 0.0   & $+0.26$ \\
     &      &        &        & 0.01  & $+0.26$ \\
     &      &        &        & 1.0   & $+0.24$ \\
     &      &        &        & 100   & $+0.04$ \\
5690 & 3.67 & $-2.5$ & $+0.5$ & 0.0   & $+0.25$ \\
     &      &        &        & 0.01  & $+0.25$ \\
     &      &        &        & 1.0   & $+0.24$ \\
     &      &        & $+1.0$ & 0.0   & $+0.27$ \\
     &      &        &        & 0.01  & $+0.27$ \\
     &      &        &        & 1.0   & $+0.26$ \\
6450 & 4.04 & $-3.0$ & $+1.0$ & 0.0   & $+0.15$ \\
     &      &        &        & 0.01  & $+0.15$ \\
     &      &        &        & 1.0   & $+0.15$ \\
\hline
\end{tabular}
\begin{list}{}{}
\item[$^{\rm a}$] The solar O abundance is here assumed to be 
${\rm log} \epsilon_{\rm O} = 8.90$
\item[$^{\rm b}$] The factor which is multiplied to the Drawin (1968) recipe
for the collisional cross-sections with H
\item[$^{\rm c}$] The non-LTE O abundance correction defined as
$\Delta {\rm log} \epsilon_{\rm O} = 
{\rm log} \epsilon_{\rm O}^{\rm NLTE} -
{\rm log} \epsilon_{\rm O}^{\rm LTE} $
\end{list}
\end{table}

Non-LTE calculations have been performed for four 1D model atmospheres
with the solar $T_{\rm eff}$ and log\,$g$ but different metallicities
([Fe/H]\,$=0.0$, $-1.0$, $-2.0$, and $-3.0$), as well as for selected 
metal-poor stars (e.g. HD\,140283 and G64-12) for different O abundances 
to estimate the influence of the line strength. In all cases, 
the effect of H collisions were investigated by computing the
abundance corrections for three different $x$: $0.0$, $0.01$ and $1.0$. 
The 1D models are not identical
to those adopted for the solar sequence in the 3D-1D LTE comparison
in order to strictly isolate
the metallicity dependence of the non-LTE effects. 

\begin{figure}[t]
\resizebox{\hsize}{!}{\includegraphics{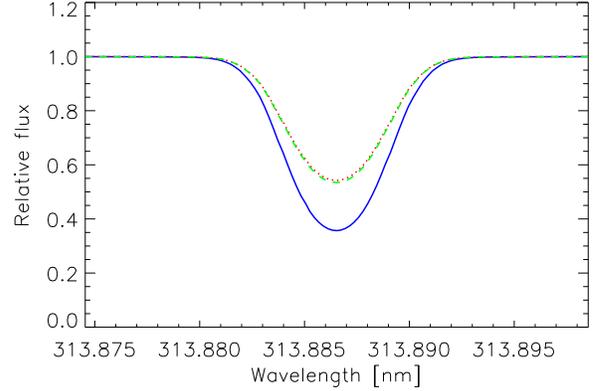}}
\caption{The 1D OH\,313.9\,nm line 
in a metal-poor ([Fe/H]\,$=-3.0$) Sun 
in LTE (solid lines), and in non-LTE with $x=1.0$ (dashed lines) 
and $x=0.0$ (dotted lines), see text for details.
Clearly the inclusion of collisions by H according to the 
Drawin (1968) formula ($x=1.0$) has a negligible impact on
the resulting line strength
}
         \label{f:nlte_prof}
\end{figure}

A summary of the results in terms of abundance corrections
is presented in Table \ref{t:nlte}.
As expected, in all cases the assumption of LTE makes the lines stronger,
and therefore that LTE analyses underestimate the O abundances.
The difference in terms of line strengths is quite large 
(Fig. \ref{f:nlte_prof}) and therefore
relatively large non-LTE corrections are obtained, $\simeq 0.2$\,dex. 
The predicted 1D non-LTE effects are almost independent on the stellar parameters.
Fortunately, the magnitude of the non-LTE corrections is only marginally dependent 
on the treatment of the H collisions provided the Drawin (1968) formula
does not greatly underestimate the collisional cross-sections ($x>10$). 
Even with $x=1.0$ the predicted non-LTE corrections differ only by $\le 0.03$\,dex 
compared to the case when neglecting the H collisions completely. In all
cases, the same results are obtained with $x=0.01$ and $x=0.0$. 
We emphasize though that it is still unclear whether the Drawin
recipe can be applied also to molecules. But at this stage there are 
no indications that the treatment of H collisions play any significant
role in the predicted non-LTE corrections.

\begin{figure}[t]
\resizebox{\hsize}{!}{\includegraphics{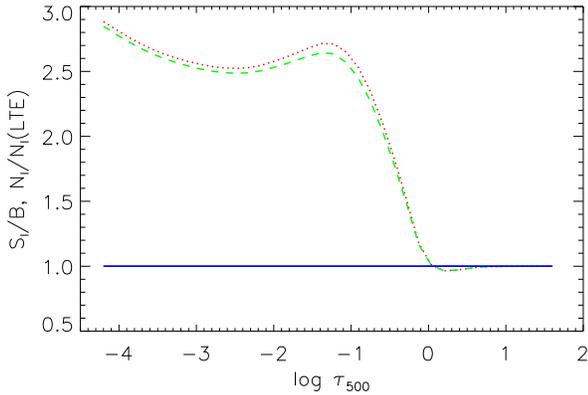}}
\caption{The ratio of the line source function with the Planck function
$S_{\rm l}/B_\lambda(T)$ for the two cases $x=1.0$ (dashed lines)
and $x=0.0$ (dotted lines) in 
a metal-poor ([Fe/H]\,$=-3.0$) Sun,
see text for details. Also shown are the departure coefficients 
$\beta = N_{\rm l}/N_{\rm l}^{\rm LTE}$ for the lower level of the OH 
transition for $x=1.0$ and $x=0.0$ (solid lines), 
emphasizing that the non-LTE effect is purely
due to scattering and not changes in the line opacity
}
         \label{f:nlte_sl}
\end{figure}

The departures from LTE in this two-level approach are purely due to
scattering in the line, in agreement with the prediction by
Hinkle \& Lambert (1975) (Fig. \ref{f:nlte_sl}).  
Since $\bar{J_\lambda}/B_\lambda(T) > 1$ at the relevant
atmospheric layers for the studied model atmospheres, the line source function
$S_{\rm l}$ exceeds the LTE value. This effect is more pronounced in
the metal-poor models but its influence on the abundance corrections
is partly balanced by the weaker lines at lower metallicities. 
Since the vast majority of the OH molecules will be in the ground
electronic state and the vibrational and rotational levels within
electronic states are closely coupled through collisions, 
the departure coefficient 
$\beta = N_{\rm l}/N_{\rm l}^{\rm LTE}$ for the lower level
will be very close to one and thus no non-LTE effects due to differences
in line opacity emerge, as evident from Fig. \ref{f:nlte_sl}.

In spite of the significant 1D non-LTE abundance corrections, 
we find no evidence for a pronounced steepening of the [O/Fe] trend.
However, we caution that this conclusion can not automatically
be extrapolated to the 3D case, in which the
departures from LTE may be more severe in metal-poor
stars since the steep temperature gradients  
may be more prone to scattering
effects than in 1D model atmospheres. Needless to say, an investigation
of the non-LTE behaviour in 3D model atmospheres
as a function of metallicity has very high priority.

\subsection{Summary of uncertainties: 3D or 1D model atmospheres? 
\label{s:summary}}

Considering that the results presented here are among the first
investigations of the impact of the new generation of 3D hydrodynamical
model atmospheres on stellar spectroscopy, it is therefore in order
to ask if indeed these models are more realistic than previously
used classical 1D model atmospheres. 
Sofar the predictions from the {\em ab-initio}
3D models have been very successfully
confronted with detailed observational constraints for 
in particular the Sun.
These comparisons include such disparate tests as granulation
topology and flow field (Stein \& Nordlund 1998), helioseismology
(Rosenthal et al. 1999), intensity brightness contrast
(Stein \& Nordlund 1998; Asplund et al. 2000a), 
flux distribution and limb-darkening (Asplund et al. 1999b)
and spectral line shapes, shifts and asymmetries (Asplund et al. 2000b).
No doubt the current surface convection simulations for the Sun have
a very high degree of realism. In sharp contrast theoretical 1D model
atmospheres fail in regards to most, if not all, of the
above-mentioned tests.
Recently similar 3D models have been used for studies of line
asymmetries in Procyon (Allende Prieto et al. 2001) and the metal-poor
halo star HD\,140283 
(Allende Prieto et al. 1999a; Asplund et al., in preparation) 
with very satisfactory outcomes.

In view of the detailed discussion given in Sect. \ref{s:errors} 
on possible remaining systematic errors in the 3D analysis, 
the reader may get
the impression that the here presented results are rather uncertain.
It may therefore be in order to point out that all of these possible
effects also apply equally well to any study based on 1D model atmospheres,
besides the errors introduced by the assumption of 
hydrostatic equilibrium and by treating convection through the mixing 
length theory (or a close relative thereof), which is known to 
be a poor representation of stellar convection.  
However, at this stage it is probably premature to conclude 
that 3D model atmospheres are indeed superior to classical 1D models.
It is therefore of utmost importance now to carry out the same arsenal of
tests which previously has been undertaken with 1D model atmospheres 
(flux distribution, limb-darkening, colours, H-lines etc) as well
as additional ones now possible 
(line asymmetries and shifts, asteroseismology),
in particular for metal-poor stars. 
This is even more true in light of the fact that 1D model atmospheres
often fail the very same tests.

\section{Comparing previous OH, O\,{\sc i} and [O\,{\sc i}] results
\label{s:odiagnostics}}

Even if the present article does not deal directly with observations,
in this section we will nevertheless discuss some of the recent
analyses of various O diagnostics as it has bearing on our findings
and may give clues to the existence or not of the large granulation
corrections for OH in metal-poor stars described in Sect. \ref{s:oh}. 

\begin{figure}[t]
\resizebox{\hsize}{!}{\includegraphics{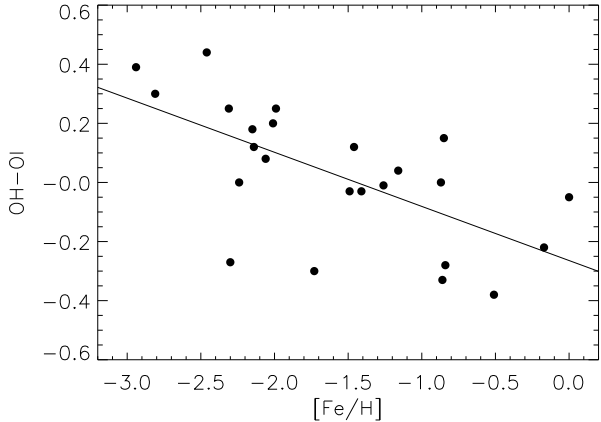}}
\caption{The difference between the OH and 
O\,{\sc i}-based 1D abundances of Boesgaard et al. (1999) on the King 
(1993) $T_{\rm eff}$-scale, which reveals a pronounced 
metallicity-dependent trend (slope $-0.18$). This behaviour is in
qualitative agreement with the 3D calculations presented in Sect. \ref{s:oh}
}
         \label{f:ohvsoi}
\end{figure}

A major argument for the monotonic linear trend in [O/Fe] with
metallicity derived from the UV OH lines comes from the claimed
good agreement with the O\,{\sc i} triplet results
(Israelian et al. 1998, 2001; Boesgaard et al. 1999). Since much smaller
granulation corrections are expected for the triplet than for OH
(Asplund 2001) this would seem to contradict the findings in Sect. \ref{s:oh}.
However, the consonance between the OH and O\,{\sc i} results
is not as unambigous when examining some of
the published analyses in detail.
Fig. \ref{f:ohvsoi} shows the difference between the OH-based and 
O\,{\sc i}-based abundances of Boesgaard et al. (1999) on the King (1993)
$T_{\rm eff}$-scale; with the lower Carney (1983) scale the correlation
is slightly less pronounced but clearly present 
(slope~$=-0.13$ instead of $-0.18$).
The absolute value for the abundance differences is here less important than
the existence of the metallicity-trend due to the uncertainty introduced by
the choice of $T_{\rm eff}$-scale and the neglect of non-LTE effects for
O\,{\sc i}.
This divergent behaviour is very close to the expected according to 
the results of Sect. \ref{s:oh} and Asplund (2001), 
which we interpret as a qualitative argument for
the aptness of the 3D calculations. 
In view of this, the good general agreement found by Israelian et al. (1998)
for nine stars using the O\,{\sc i} equivalent widths of Tomkin et al. (1992)
pose a perplexing problem. 
Clearly a larger stellar sample with simultaneous analyses of the OH and
O\,{\sc i} lines would be very helpful in this context.
One should also explore
possible differences in for example C abundances derived from C\,{\sc i} 
and CH lines
at low metallicities. Preliminary calculations reveal significant 
but smaller differences between 1D and 3D analyses of CH lines
compared with for OH lines ($\approx 0.3$\,dex instead of 
$\approx 0.6$\,dex for OH at [Fe/H]\,$=-3.0$).

The conventional [O/Fe] plateau indicated by the [O\,{\sc i}] line has
recently received support from the IR OH vibrational-rotational lines
(Balachandran et al. 2001; Melendez et al. 2001). 
As discussed in Sect. \ref{s:oh}, this could be interpreted as the
here presented granulation corrections for the UV OH lines in metal-poor
stars are overestimated, since for the same star the IR and UV lines
should to first order be affected similarly (with the possible exception
of non-LTE effects in the line formation, cf. Sect. \ref{s:nlte}).
A more likely explanation in our opinion, however, is to be found in
the differences in stellar parameters, in particular $T_{\rm eff}$, for
the sofar investigated UV and IR targets. As evident from Table \ref{t:oh}
and Fig. \ref{f:OFe}, the magnitude of the OH granulation corrections
depend on $T_{\rm eff}$ in the sense that larger effects are present for
higher temperatures. 
Although dangerous to extrapolate, we speculate that significantly smaller
OH abundance corrections than found here 
will be obtained for $T_{\rm eff} \la 5000$\,K,
which is typical for the metal-poor stars with detected IR OH features.
The physical reason for the dependence on $T_{\rm eff}$ is the strength 
of the coupling between the gas and radiation field: for lower $T_{\rm eff}$,
more and stronger spectral lines are available which contribute additional
radiative heating, keeping the gas temperature closer to the
radiative equilibrium value (Asplund et al., in preparation).
No 3D model atmospheres are yet available for these lower $T_{\rm eff}$ 
which prevent verification of our hypothesis but such convection 
simulations are currently being constructed.

Due to the inconspicous nature of the [O\,{\sc i}] line (Nissen et al. 2001),
no direct comparison with the UV OH results has sofar 
been possible for metal-poor dwarfs with [Fe/H]~$<-2$.
Recently, Fulbright \& Kraft (1999) have considered the [O\,{\sc i}] line
in the two metal-poor ([Fe/H]~$\la-2.1$) 
subgiants BD$+23\degr 3130$ and BD$+37\degr 1458$ 
($T_{\rm eff} = 5130$\,K and 5260\,K, respectively), which have been
re-analysed by Israelian et al. (2001). 
The UV OH lines suggest [O/Fe]~$=0.60$ and $0.50$, respectively, in the two
stars, i.e. in significantly better agreement with the traditional 
[O/Fe] plateau compared with the linear trend seemingly
implied by the OH lines in the hotter stars. 
As for the IR analyses, we conjecture that at
$T_{\rm eff} \simeq 5200$\,K significantly smaller granulation corrections
than those given in Sect. \ref{s:oh} will be deduced,
leaving the plateau-like [O/Fe] ratios largely unchallenged.  
Excellent agreement between [O\,{\sc i}] and OH is found for both
stars: the forbidden line 
gives [O/Fe]~$=0.62$ and $0.52$, respectively.   
The latter values come from our own analysis using 1D {\sc marcs} 
model atmospheres with the stellar parameters given in
Israelian et al. (2001). The equivalent widths for [O\,{\sc i}]
and Fe\,{\sc ii} were taken from Fulbright \& Kraft except for [O\,{\sc i}]
in BD$+23\degr 3130$ for
which we relied on the VLT-measurement ($W_\lambda = 0.15$\,pm) 
by Cayrel (2001). 
This [O/Fe] estimate for
BD$+23\degr 3130$ is for unknown reason significantly lower
than reported in Israelian et al. (2001), [O/Fe]~$=0.82$, while
in perfect agreement with the findings of 
Balachandran (2001, private communication) using Kurucz (1993) model
atmospheres but otherwise identical input.

\section{Concluding remarks 
\label{s:conclusions}}

The low atmospheric temperatures encountered in 
the new generation of 3D hydrodynamical model atmospheres
(Asplund et al. 1999a, 2000a,b, Allende Prieto et al. 2001;
Asplund et al., in preparation) compared with classical 
1D hydrostatic model atmospheres for 
metal-poor stars have a profound impact on
the OH line formation, as described in
Sect. \ref{s:oh}. As a consequence, a possible 
severe systematic error in recent
1D LTE analyses of OH lines in metal-poor stars
(Israelian et al. 1998, 2001; Boesgaard et al. 1999) has been exposed, 
making the case for a monotonic increase in [O/Fe] towards lower metallicities
less convincing.
Taken at face value, our 3D LTE OH analysis results in [O/Fe] values
in rough agreement with the conventional [O/Fe] plateau for
[Fe/H]~$\la -1$ indicated by the [O\,{\sc i}] lines (Fig. \ref{f:OFe_average}).
We emphasize though that this apparant concordance should not
be taken too literally in view of the preliminary nature of
our 3D calculations and that no star-by-star comparison has been
made with observations. 
By investigating possible systematic errors for the 3D LTE results
it has been found, however, that the conclusion 
of large granulation effects on the OH lines appears reasonably robust. 
For example, $T_{\rm eff}$-calibrations using 3D model atmospheres
should not differ significantly ($\Delta T_{\rm eff} \la 20$\,K)
from previous 1D calibrations provided
they are based on IRFM; the same is not necessarily true for alternative
calibrations using Balmer lines or colours. 
Similarly, estimates of log\,$g$ and [Fe/H] should remain essentially
unaltered when relying on Hipparcos parallaxes and Fe\,{\sc ii} lines. 
The major remaining uncertainties in the 3D analysis appear to be the 
assumptions of LTE for the molecular equilibrium and in the line formation.
Although not yet investigated for 3D model atmospheres, it is possible
that such departures from LTE may indeed steepen the [O/Fe] trend once again
but unlikely as much as the original 1D LTE case.
On the other hand, departures from LTE for Fe\,{\sc i} may diminish 
the slope further.
A final verdict on this issue must therefore await detailed and improved 
non-LTE calculations. But we note in the meantime that even in the 
presence of possible non-LTE effects for OH and Fe\,{\sc i}, [O/Fe]
ratios will be less affected since both the O and Fe LTE abundances
will tend to be underestimated.

The purpose of the present paper has not been to advocate a specific [O/Fe]
trend with metallicity since our investigation has been limited to a 
differential 3D-1D comparison without involving observational confrontation. 
Nevertheless, our findings will likely fuel the long-standing
debate on the O abundances in metal-poor stars.
As already stated several times, it would be premature to conclude from
our analysis that the monotonic linear trend in [O/Fe] claimed by
Israelian et al. (1998, 2001) and Boesgaard et al. (1999)
must now be abandoned, even if the arguments for continuously increasing
[O/Fe] ratios towards lower metallicities appear much weaker. 
It should be remembered that many, albeit not all, studies of the
O\,{\sc i} triplet at 777\,nm in dwarfs and subgiants
find [O/Fe] values systematically higher
than those given by the [O\,{\sc i}] line in metal-poor giants and subgiants.
The triplet is unfortunately sensitive to the adopted $T_{\rm eff}$-scale
and departures from LTE, which deserve very careful treatment. 
For the moment, the most reliable [O/Fe] ratios still appear to come from
the [O\,{\sc i}] line, provided that very high $S/N$ spectra are
utilised and the metallicities are estimated from Fe\,{\sc ii} lines
(Nissen et al. 2001; Lambert 2001).

To ignore the existing systematic errors in traditional 1D LTE analyses
of the UV OH lines in the hope that departures from LTE in 
the molecule formation
and radiative transfer will conspire to exactly compensate the effects
of the low atmospheric temperatures in 3D is certainly fraught with danger
and furthermore most likely misleading.
As demonstrated above, 
in order to derive reliable oxygen abundances in very metal-poor stars,
unfortunately there is no easier escape route but to perform 
time-dependent non-LTE calculations in 3D model atmospheres for OH. 
Fortunately, recent improvements in the analysis of stellar spectra
ensure such studies in fact being tractable tasks for the near future,
which should help remove many of the lingering uncertainties in
the derived oxygen abundances and
perhaps finally settle the long-standing debate on [O/Fe]  
in metal-poor stars.

\begin{acknowledgements}
It is a pleasure to acknowledge many stimulating discussions with 
S. Balachandran, B. Gustafsson, T. Karlsson,
D. Kiselman, D.L. Lambert, P.E. Nissen, F. Primas and N. Ryde
related to the determinations of stellar oxygen abundances and their
associated intricacies.
The efforts by the two referees
are much appreciated.
The present work has been supported by the Swedish Natural Science
Foundation (grant NFR F990/1999), by the Royal Swedish Academy
of Sciences and by the Nordic Optical Telescope
through a PhD stipend to AEGP.
MA is grateful for IAU travel grants to attend the IAU Symposium 198
in Natal, Brazil, and the IAU General Assembly in Manchester, UK.
\end{acknowledgements}

\end{document}